\documentclass[default]{sn-jnl}

\usepackage{graphicx}%
\usepackage{multirow}%
\usepackage{amsmath,amssymb,amsfonts}%
\usepackage{amsthm}%
\usepackage{bm}
\usepackage{caption}
\usepackage{subcaption}
\usepackage{mathrsfs}%
\usepackage[title]{appendix}%
\usepackage{xcolor}%
\usepackage{textcomp}%
\usepackage{manyfoot}%
\usepackage{booktabs}%
\usepackage[ruled]{algorithm2e}
\usepackage{algorithmicx}%
\usepackage{algpseudocode}%
\usepackage{listings}%
\usepackage{comment}
\usepackage{amsmath}
\usepackage{graphicx}
\usepackage{amssymb} 
\usepackage{amsthm}
\usepackage{amsmath}
\usepackage{caption}
\usepackage{tabularx}
\usepackage{bm}
\usepackage{float}
\usepackage{placeins}
\setcitestyle{authoryear,open={(},close={)}}

\theoremstyle{thmstyleone}%

\theoremstyle{thmstyletwo}%

\theoremstyle{thmstylethree}%

\begin{document}

\title[Article Title]{ Bayesian Tensor Modeling for Image-based Classification of Alzheimer's Disease}

\author*[1]{\fnm{Rongke} \sur{Lyu}}\email{rl77@rice.edu}

\author[1]{\fnm{Marina} \sur{Vannucci}}

\author[2]{\fnm{Suprateek} \sur{Kundu}}

\affil[1]{\orgdiv{Department of Statistics}, \orgname{Rice University}, \orgaddress{\city{Houston}, \state{TX}, \country{United States}}}

\affil[2]{\orgdiv{Department of Biostatistics}, \orgname{MD Anderson Cancer Center}, \orgaddress{\city{Houston}, \state{TX}, \country{United States}}}

\abstract{Tensor-based representations are being increasingly used to represent complex data types such as imaging data, due to their appealing properties such as dimension reduction and the preservation of spatial information. Recently, there is a growing literature on using Bayesian scalar-on-tensor regression techniques that use tensor-based representations for high-dimensional and spatially distributed covariates to predict continuous outcomes. However surprisingly, there is limited development on corresponding Bayesian classification methods relying on tensor-valued covariates. Standard approaches that vectorize the image are not desirable due to the loss of spatial structure, and alternate methods that use extracted features from the image in the predictive model may suffer from information loss. We propose a novel data augmentation-based Bayesian classification approach relying on tensor-valued covariates, with a focus on imaging predictors. We propose two data augmentation schemes, one resulting in a support vector machine (SVM) type of classifier, and another yielding a logistic regression classifier. While both types of classifiers have been proposed independently in literature, our contribution is to extend such existing methodology to accommodate high-dimensional tensor valued predictors that involve low rank decompositions of the coefficient matrix while preserving the spatial information in the image. An efficient Markov chain Monte Carlo (MCMC) algorithm is developed for implementing these methods. Simulation studies show significant improvements in classification accuracy and parameter estimation compared to routinely used classification methods. We further illustrate our method in a neuroimaging application using cortical thickness MRI data from Alzheimer’s Disease Neuroimaging Initiative, with results displaying better classification accuracy throughout several classification tasks.

}


\maketitle

\section{Introduction}\label{Intro}
Neuroimaging studies stand as a cornerstone in contemporary neuroscience, fundamentally transforming our comprehension of the intricate structure and function of the brain. These non-invasive visualization techniques have not only enriched our understanding of neurological disorders but have also pioneered new frontiers in mental health research. Within the realm of risk prediction, neuroimaging studies have emerged as an invaluable tool for identifying individuals susceptible to neurological and psychiatric conditions. By discerning subtle abnormalities in brain structure and connectivity, researchers can now predict the risk of mental disorders (such as Alzheimer's disease or dementia) with greater precision. This early identification facilitates timely intervention and treatment, potentially offering an opportunity to provide improved health outcomes and quality of life related to these disorders. 

For example, neuroimaging studies have transformed the field of Alzheimer's disease (AD) research, providing invaluable insights into the pathological mechanisms underlying this devastating neurodegenerative disorder \citep{chouliaras2023use}. By visualizing the brain's structural and functional changes, neuroimaging techniques have enabled researchers to track the progression of AD, identify early signs of the disorder, and differentiate it from other causes of dementia. In particular, structural neuroimaging studies involving magnetic resonance imaging (MRI), have revealed the intricate patterns of atrophy in spatially distributed brain regions involved in memory and cognition that is a hallmark of AD \citep{frenzel2020biomarker}. Neuroimaging features such as brain volume or cortical thickness that are derived from MRI scans can be used to provide quantitative assessments of disease severity and monitor disease progression over time. Importantly, such neuroimaging features can be embedded in machine learning algorithms in order to perform risk prediction or early detection in AD. 
However, several major challenges are encountered when analyzing neuroimaging data. For example, the brain imaging data is spatially dependent, high-dimensional and noisy, and it is often unclear how to identify suitable neurobiological markers for the mental disorder in the presence of heterogeneity. 

In order to model such complex types of imaging data emerging at a rapid pace, several statistical and machine learning approaches have been proposed. Among them, classification models using neuroimaging features have seen a rapid development \citep{rathore2017review,arbabshirani2017single,falahati2014multivariate}. These approaches typically either vectorize the image, or extract informative summary features from the image, to be used as covariates.  For example, \cite{plant2010automated} extracted the low-level-feature extraction algorithm with feature selection criterion to select most discriminating features, that are then coupled with a clustering algorithm to group spatially coherent voxels to predict Alzheimer's disease status. \cite{ben2015classification} proposed a multi-feature fusion algorithm used both extracted visual features from the hippocampal region of interest (ROI) and the quantity of cerebrospinal fluid (CSF) in the hippocampal region, and then applied a late fusion scheme to perform the binary classification of Alzheimer's disease subjects using the MRI images. Going beyond AD classification, \cite{griffis2016voxel} implemented a voxel-based Gaussian Naive Bayes Classification of ischemic stroke lesions in individual T1-weighted MRI scans, where the authors separately created two feature maps as predictor variables for missing and abnormal tissue to avoid including highly redundant information. Alternate types of shape-based image analysis that go beyond voxel-level analysis have also been proposed for prediction \citep{wu2022elastic}.

The above approaches, while useful, did not explicitly account for the spatial configuration of imaging voxels. Some exceptions include Markov random field (MRF) based methods that have been proposed in the prediction context \citep{smith2007spatial, lee2014spatial}. However, given that these are not equipped to perform dimension reduction, they may not be fully scalable to high-dimensional images with tens of thousands of voxels, and their performance in classification problems is unclear. In order to tackle the spatial information in the image in the context of multi-class classification, \cite{pan2018covariate} proposed a penalized linear discriminant analysis (LDA) model using scalar and tensor covariates. Unfortunately, there is limited, if any, literature on Bayesian classification approaches based on imaging features that account for the spatial information in the image. This is surprising, given the utility of Bayesian methods that can predict the probability of an observation belonging to two or more classes, which can be useful in the presence of measurement error or uncertainty regarding class labels in medical imaging studies \citep{morales2013predicting,behler2022multivariate}. Existing Bayesian classification approaches that use vectorized features can be not readily adapted to our problem of interest involving Bayesian image-based classification, since it ignores the spatial structure of the image resulting in information loss and potentially poor model performance. Additionally, simply vectorizing the imaging features without an appropriate lower dimensional representation also introduces the curse of dimensionality since the number of voxels in the image is typically tens of thousands. Alternate approaches that rely on first extracting lower dimensional features from the image and subsequently using these features for classification, may involve an additional layer of information loss resulting from the feature extraction step, resulting in potential loss in accuracy.

Recently, there has been a growing literature on tensor analysis in statistical modeling for imaging data that addresses some of the above concerns. \cite{guhaniyogi2017bayesian} proposed a Bayesian tensor regression with a scalar response on scalar and tensor covariates. Other tensor models include Bayesian response regression models that model the image outcome as a tensor object. \cite{guhaniyogi2021bayesian} implemented a Bayesian Tensor response on scalar regression with an application of neuronal activation detection in fMRI experiments using both tensor-valued brain images and scalar covariates. \cite{kundu2023bayesian} proposed a longitudinal Bayesian tensor response regression model for mapping neuroplasticity across longitudinal visits. \cite{billio2023bayesian} proposed a novel linear autoregressive tensor process model that introduces dynamics in linear tensor regression and allows for both tensor-valued covariates and outcomes. Under the frequentist approach, \cite{lock2018tensor} proposed a penalized tensor on tensor regression using a (L2) Ridge penalty. \cite{zhou2013tensor} proposed a tensor regression model by extending generalized linear model to include tensor-structured covariates, where the rank-\textit{R} PARAFAC decomposition is assumed on the tensor parameters, with adaptive lasso penalties applied on the tensor margins. With the exception of the penalized GLM approach in \cite{zhou2013tensor}, most existing tensor-based approaches in literature have mainly focused on linear regression models that can not be readily used for Bayesian classification.

In this article, we propose a data augmentation-based Bayesian classification approach that models binary outcomes based on imaging covariates using a tensor-based representation. We consider two different data augmentation schemes resulting in two distinct Bayesian classifiers: a support vector machine (SVM) and a logistic regression model. While these classifiers have been extensively used in literature, the focus has been on using non-structured covariates that ignore the spatial structure embedded in the image. Our specific interest is in Bayesian classification based on imaging predictors, where the images are registered across samples. Such a set-up is routinely used in neuroimaging studies. Our main contribution is to develop a Bayesian classification methodology based on high-dimensional tensor-valued predictors via low-rank decompositions of the coefficient matrix and using data augmentation. The low-rank PARAFAC decomposition assumed by the tensor model is able to preserve the spatial configuration of imaging voxels, while overcoming the challenges arising from the high dimensionality of the image that can often contain tens of thousands of voxels. This results in considerable improvements in classification accuracy, as illustrated via rigorous numerical examples. In contrast to existing feature extraction approaches that first use a tensor decomposition or alternate schemes to obtain low level features to be subsequently used in modeling \citep{sen2021predicting}, the proposed approach uses the full image as is in the classification model, but employs a low rank PARAFAC decomposition to model the high-dimensional tensor model coefficients. This ensures no information loss due to feature extraction, while simultaneously allowing for dimension reduction.
We adopt the multiway shrinkage prior from \cite{guhaniyogi2017bayesian} to model the tensor margins of the assumed rank-\textit{R} PARAFAC decomposition, which shrinks non-significant parameters to near zero while inducing a minimal shrinkage effect on the significant parameters. We develop efficient Markov chain Monte Carlo (MCMC) algorithms for posterior inference that use data augmentation techniques. Simulation studies show significant improvements in classification accuracy, parameter estimation and feature selection compared to routinely used classification methods that use vectorized images. We further illustrate the advantages under our method via a detailed neuroimaging application using voxel-wise cortical thickness features data from Alzheimer’s Disease Neuroimaging Initiative (ADNI) study, with the proposed Bayesian classifiers displaying better out-of-sample accuracy consistently throughout several classification tasks. We leverage cortical thickness as our neuroimaging feature of choice, since it is known to be a highly sensitive imaging biomarker for modeling neurodegeneration in AD \citep{weston2016presymptomatic,fjell2015development}.

The rest of the paper is organized as follows: In section \ref{Model Framework}, we propose the framework of the Bayesian tensor classification model with two types of data augmentation, specify the prior and hyperparameter choices, and list the posterior computation steps. In Section \ref{Simulation} we study the model performances through comprehensive simulation studies. In Section \ref{ADNI-section} we provide the results from the data analysis using the ADNI dataset. We conclude the manuscript with a discussion. 

\section{Methods}\label{Model Framework}

\subsection{Brief Introduction to Tensors}\label{Tensor Decomposition}
Tensor-based models have gained recognition as a promising way to model neuroimaging data, due to their multifold advantages. Tensors naturally inherit a multidimensional structure to represent complex data structures such as spatial features of a brain region. Additionally, tensor-based techniques achieve dimension reduction, which is particularly useful with neuroimaging data to tackle the challenges of $p>>n$ in statistical modeling. A tensor is a multi-dimensional array, with the order being the number of its dimensions. For example, a one-way or first-order tensor is a vector, and a second-order tensor is a matrix. A \textit{fiber} resembles the idea of matrix rows and columns in higher dimensionality, which is obtained by fixing every dimension of a tensor except one. Similarly, a \textit{slice} is defined by fixing every order of the tensor except two. Tensor decomposition is a mathematical technique that expresses a high dimensional tensor into the combination of lower dimensional factors. One type of tensor decomposition is the Tucker decomposition \citep{kolda2009tensor}, which decomposes a tensor into a core tensor and a set of matrices, one along each mode. It can be denoted as follows:

\begin{align}\label{eq:tucker}
    \boldsymbol{B}&=\Lambda \times_1 A \times_2 B \times_3 \cdots \times_D D \nonumber \\
    &= \sum_{r_1=1}^{R_1}  \cdots \sum_{r_D=1}^{R_D} \lambda_{r_1, \ldots, r_D} \boldsymbol{a}^{\left(r_1\right)} \circ  \cdots \circ \boldsymbol{d}^{\left(r_D\right)},
\end{align}

\noindent where $\Lambda$ is the core tensor, and $A$, $B$, $D$ are the factor matrices. The PARAFAC decomposition is a special case of the Tucker decomposition, where the core tensor $\Lambda$ is restricted to be diagonal, and $R_1 = R_2 =\ldots=R_D = R$. The rank-\textit{R} PARAFAC model can then be expressed as

\begin{align}\label{eq:parafac}
    \boldsymbol{B}=\sum_{r=1}^R \boldsymbol{\beta}_1^{(r)} \circ \cdots \circ \boldsymbol{\beta}_D^{(r)},
\end{align}

\noindent where $\boldsymbol{\beta}_1,\ldots, \boldsymbol{\beta}_D$, known as tensor margins, are vectors of length $p_1,\ldots,p_D$, and where $\boldsymbol{\beta}_1 \circ \cdots \circ \boldsymbol{\beta}_D$ is a D-way outer product of dimension $p_1 \times p_2 \times\ldots\times p_D$. It is essential to recognize that tensor margins can only be uniquely identified up to a permutation and a multiplicative constant unless we introduce additional constraints. However, the lack of identifiability in tensor margins does not create any complications for our scenario. This is because the tensor product is fully identifiable, which suffices for our primary objective of estimating coefficients. Consequently, we refrain from imposing extra identifiability conditions on the tensor margins, aligning with the principles in the Bayesian tensor modeling literature \citep{guhaniyogi2020bayesian}. Furthermore, the PARAFAC decomposition dramatically reduces the number of coefficients from $p_1 \times \ldots \times p_D$ to $R(p_1 + \ldots + p_D)$, which grows linearly with the tensor rank $R$ and results in significant dimension reduction. The appropriate tensor rank can vary depending on the specific application context and can be chosen using a goodness-of-fit approach.

Prior to applying the tensor model, the image's voxels are transformed onto a regularly spaced grid, making them more suitable for a tensor-based approach. This mapping conserves the spatial arrangements of the voxels, offering notable advantages over a univariate voxel-wise analysis or a multivariable analysis that vectorizes the voxels without considering their spatial arrangements. While the grid mapping might not preserve exact spatial distances between voxels, this has limited impact, as it can still capture correlations between neighboring elements in the tensor margins. Moreover, the tensor construction has the advantageous ability to estimate voxel-specific coefficients by leveraging information from neighboring voxels through the estimation of tensor margins with their inherent low-rank structure. This feature results in brain maps which are more consistent and robust to missing voxels and image noise. Additionally, it can be conveniently used for reliable imputation of imaging features related to missing voxels. In contrast, voxel-wise analysis lacks the capacity to share information among neighboring voxels, treating voxel coefficients as independent entities regardless of their spatial arrangement. For more details on the advantages and characteristics of Bayesian tensor models, we refer readers to \cite{kundu2023bayesian}.

\subsection{Loss-function Based Classification} 

A loss function is a mathematical tool to quantify the difference between predicted values under a model and the actual observed data values. Loss functions are often used for finding optimal estimates of model parameters in machine learning and statistical modeling tasks for regression, classification, and more. In the Bayesian paradigm, loss functions translate to different types of likelihood that are combined with additional priors on the model parameters embedded in the loss function to obtain posterior distributions that are subsequently used for estimation and uncertainty quantification. Although we motivate our approach by drawing connections with loss functions, we note that our work is distinct compared to decision theoretic Bayesian approaches that use loss functions as a post-processing step after Markov chain Monte Carlo (MCMC) to derive optimal estimates. See, for example, \cite{hahn2015decoupling} and \cite{kundu2019efficient}. In this article we focus, in particular, on two types of commonly used loss functions: the hinge loss represented as the support vector machine classifier, and the logistic regression loss, both of which employ high-dimensional images as covariates for classification. 

\begin{figure}[h]%
\centering
\includegraphics[width=1\textwidth]{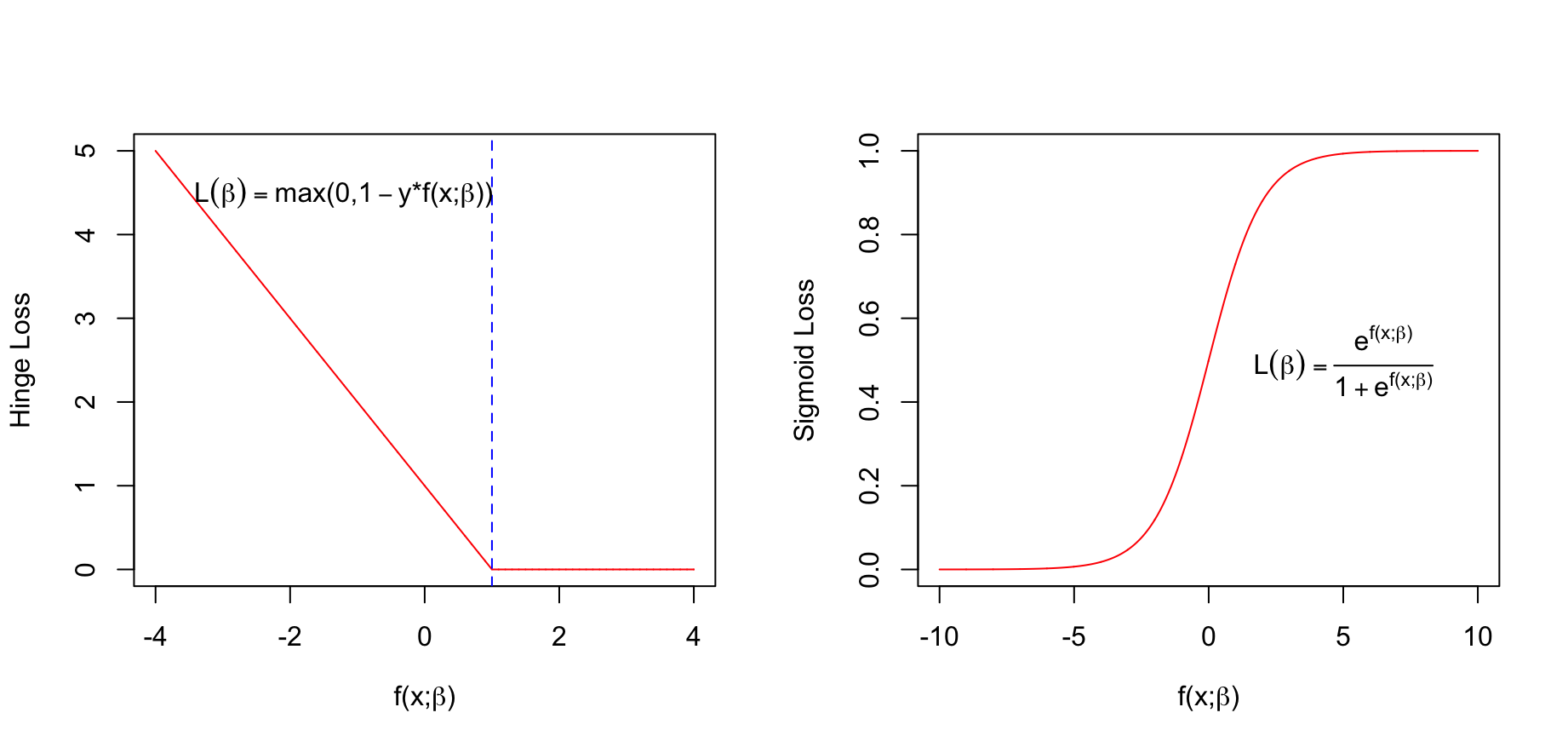}
\caption{From left to right: hinge loss and sigmoid loss function}\label{fig1}
\end{figure}

\vskip 12pt

{\noindent \underline{\bf Support vector machine (SVM):}} SVMs play a pivotal role in classification tasks, and their significance stems from their ability to handle complex decision boundaries with remarkable efficiency. SVMs work by finding an optimal hyperplane that separates the data points into two classes. This hyperplane is chosen to maximize the margin, which is the distance between the hyperplane and the closest data points. This helps to ensure that the SVM model is generalizable to new data, and protects against overfitting. SVMs are well-suited for both linear and non-linear classification, via using suitable kernel functions. This versatility makes SVMs applicable across various domains, from image recognition and natural language processing to bioinformatics - see \cite{cervantes2020comprehensive} for a review. Their robust performance, ability to manage high-dimensional data, and capacity to handle intricate relationships between features underscore their importance in tackling diverse and challenging classification problems in machine learning.

Most SVM-based classifiers rely on point estimates with penalized approaches to tackle high-dimensional covariates \citep{peng2016review,dedieu2019error}. The SVM classifier uses the hinge loss function that takes the form 
\begin{eqnarray}
\mathcal{L}(y\mid \beta) = \frac{1}{\sigma^2} max(1-y f({\bf x}; {\bm \beta}),0) + R, 
\end{eqnarray}
where $y\in \{-1,1 \}$ is the binary outcome, $f(\cdot)$ is a linear or non-linear function of (potentially high-dimensional) covariates ${\bf x}$, with corresponding unknown parameters ${\bm \beta}$ that need to be estimated from the data, and additional tuning parameters $\sigma^2$. See Figure \ref{fig1} for a visualization of the hinge loss function. Recently, \cite{ma2022highdimensional} generalized the hinge loss to a smooth hinge loss and derived provably flexible estimators in the presence of noisy high-dimensional covariates in the SVM framework. Bayesian approaches for SVM based on a pseudo-likelihood approach were proposed originally by \cite{polson2011data} and subsequently adopted for biomedical applications in \cite{sun2018knowledge}.  In particular, the pseudo-likelihood can be represented as a location-scale mixture of normals with latent variable $\rho$ as 
\begin{align}\label{liklihood-svm}
 && L = \prod_{i=1}^n L_i(y_i\mid {\bf x}_i, {\bm \beta}, {\sigma^2}) = \prod_{i=1}^n \big\{ \frac{1}{\sigma^2}\exp \{-\frac{2}{\sigma^2} \max (1-y_i f({\bf x}; {\bm \beta}), 0)\} \big\} \nonumber \\
 &&=\int_0^{\infty} \prod_{i=1}^n \frac{1}{{\sigma} \sqrt{2 \pi \rho_i}} \exp (- \frac{(1+\rho_i-y_i f({\bf x}; {\bm \beta}))^2}{2\rho_i {\sigma^2}}) d \rho_i,
\end{align}
where $L_i$ represents the contribution corresponding to the $i$th sample. The above representation essentially uses data augmentation techniques, introducing a latent variable $\rho$ which, when marginalized over, gives back the hinge loss function. Such a latent variable representation enables an efficient Gibbs sampler for posterior inference, which will be described in Section \ref{Posterior Computation} below.

\vskip 12pt

{\noindent \underline{\bf Logistic regression classifier:}} Logistic regression is a powerful and versatile machine learning algorithm that is widely used in diverse applications, including medical diagnosis, fraud detection, customer segmentation, marketing campaigns, and recommendation systems. It is characterized by simplicity in the implementation and interpretability of the results, as the logistic regression coefficients can be understood in terms of odd-ratios. In order to make this model scalable for high dimensional biomedical applications, penalized versions of logistic regression models have been proposed  \citep{doerken2019penalized,devika2016analysis}. 

The logistic loss function specifies a sigmoid type loss that takes the analytical form 
\begin{eqnarray}
\mathcal{L}(y=1\mid {\bm \beta}) = \exp\{ f({\bf x}; {\bm \beta})\}/\big(1 + \exp\{ f({\bf x}; {\bm \beta})\} \big),
\end{eqnarray}
where $f({\bf x}; {\bm \beta})$ represents the contribution of the covariates to the logistic loss that is quantified via the unknown parameters ${\bm \beta}$, to be estimated from the data. See Figure \ref{fig1} for a visualization of the sigmoid loss function. The binary outcome variables $y\in \{0,1 \}$ are assumed to follow a Bernoulli distribution, with the corresponding probability function resembling the form $\mathcal{L}(y=1\mid {\bm \beta})$. Typically $f(\cdot)$ represents linear functions of covariates that facilitate straightforward interpretations for the model parameters, although non-linear logistic regression models have also been proposed \citep{tokdar2007posterior}. In the Bayesian paradigm, model fitting and inference for logistic regression is often achieved using Polya-Gamma latent variables \citep{polson2013bayesian}. A random variable $X$ has a Polya–Gamma distribution with parameters $b>0$ and $c \in \mathcal{R}$, denoted as $X \sim PG(b,c)$ if
\begin{align}
    X \stackrel{D}{=} \frac{1}{2 \pi^2} \sum_{k=1}^{\infty} \frac{g_k}{(k-1 / 2)^2+c^2 /\left(4 \pi^2\right)},
\end{align}
where $g_k$ follows a Gamma distribution $Ga(b,1)$. The introduction of the Polya-Gamma latent variable allows one to represent the binomial likelihood as mixtures of Gaussians. It can be shown that the logistic loss function can be recovered by marginalizing out the latent Polya-Gamma latent variable using the following relationship 
\begin{align}
    \label{likelihood-lr} \frac{\big(e^{f(\cdot)})\big)^y}{\big(1+e^{f(\cdot)}\big)^b}=2^{-b} e^{\kappa \psi} \int_0^{\infty} e^{-\omega \psi^2 / 2} p(\omega) d \omega, \mbox{ } b>0, \kappa = y-b/2, 
\end{align}
where $\omega \sim PG(b,0)$, and $f({\bf x};{\bm\beta}) $ is a linear predictor in most cases. The above identity facilitates conjugate updates under a Gaussian prior, conditional upon the latent Polya-Gamma variable, as detailed in \cite{polson2013bayesian}. The full data augmented likelihood is given by the following expression:
\begin{align}
  L = \prod_{i=1}^n  \frac{\left(e^{f_i}\right)^{y_i}}{\left(1+e^{f_i}\right)}= \prod_{i=1}^n  2^{-1} e^{\kappa_i \psi_i} \int_0^{\infty} e^{-\omega_i f_i^2 / 2} p(\omega_i) d \omega_i,
\end{align}
where $\kappa_i = y_i - 1/2$, $b=1$, $\omega_i \sim PG(1,0)$.

\subsection{Priors}\label{prior}

For our implementation, we consider the widely used linear predictor using the relationship 
\begin{eqnarray}
f_i = \langle \boldsymbol{X}_i, \boldsymbol{B}\rangle + \boldsymbol{z}_i'{\bm \gamma},
\end{eqnarray}
where $\boldsymbol{X}_i$ and $\boldsymbol{z}_i$ denote the imaging predictors and the supplemental (eg: demographic/clinical) features, respectively, for the $i$th sample, $\langle \cdot, \cdot \rangle$ denotes the inner product operator, $\boldsymbol{B}$ denotes the tensor-valued coefficient matrix that quantifies the effect of the image on the classification model, and ${\bm \gamma}$ is a vector of dimension $p_z+1$ capturing the effects of the supplemental covariates. Furthermore, we assume $\boldsymbol{B} \in \otimes^D_{j = 1} \mathbb{R}^{p_j}$, that is modeled under a PARAFAC decomposition as in (\ref{eq:parafac}).

For the prior choice on the tensor margins, we adopt the multiway Dirichlet generalized double Pareto (M-DGDP) prior from \cite{guhaniyogi2017bayesian}, which shrinks small coefficients towards zero while minimizing shrinkage of large coefficients. The prior can be expressed in hierarchical form on the tensor coefficient margins $\boldsymbol{\beta}_j^{(r)}$, $j = 1,\ldots,D$ and $r = 1,\ldots,R$  as:
\begin{align}
\label{GDPprior}
    \boldsymbol{\beta}_j^{(r)} \sim \mathrm{N}\left(0,\left(\phi_r \tau\right) \boldsymbol{W}_{j r}\right), \mbox{ } w_{j r, k} \sim \operatorname{Exp}\left(\lambda_{j r}^2 / 2\right),  
\end{align}
where $\tau \sim Ga(a_{\tau}, b_{\tau})$ is a global scale parameter, $\Phi=\left(\phi_1, \ldots, \phi_R\right)  $ follows a Dirichlet distribution that encourages shrinkage to lower rank in the assumed PARAFAC decomposition with $\left(\phi_1, \ldots, \phi_R\right) \sim Dirichlet\left(\alpha_1, \ldots, \alpha_R\right)$, and where $\boldsymbol{W}_{j r}= \mbox{Diag}(w_{j r, 1}, \ldots, w_{j r, p_j})$ are scale parameters that are margin-specific for each element and modeled under an Exponential distribution as $w_{jr,k}\sim Exp(\lambda^2_{jr}/2)$ with $\lambda_{jr}$ unknown and modeled as $\lambda_{j r} \sim \operatorname{Ga}\left(a_\lambda, b_\lambda\right)$. Additionally, one can obtain 
\begin{equation}
\beta_{j, k}^{(r)} \mid \lambda_{j r}, \phi_r, \tau \stackrel{\mathrm{iid}}{\sim} \mathrm{DE}\left(\lambda_{j r} / \sqrt{\phi_r \tau}\right), 1 \leq k \leq p_j,
\end{equation}
after marginalizing out the scale parameters $w_{j r, k}$, that is, prior \eqref{GDPprior} induces a GDP prior on the individual margin coefficients which in turn has the form of an adaptive Lasso penalty as in \cite{armagan2013generalized}. Overall, the flexibility in estimating $\boldsymbol{B}_r=\left\{\boldsymbol{\beta}_j^{(r)}; 1 \leq j \leq D\right\}$ is accommodated by component-specific scaling parameter $w_{jr,k}$ and common rate parameter $\lambda_{jr}$, which shares information between margin elements and encourages shrinkage at the local scale. We complete the prior specification by assuming a $N(0, \Sigma_{0\gamma})$ as the prior on $\mathbf{\gamma}$.

\subsection{MCMC Algorithms for Posterior Inference}
\label{Posterior Computation}
For posterior inference, we implemented efficient MCMC algorithms that take advantage of data augmentation techniques. Algorithm \ref{Alg-SVM} outlines the MCMC updates for the proposed Bayesian Tensor SVM model (BT-SVM), and Algorithm \ref{Alg-LR} illustrates the updating scheme for the the proposed Bayesian Tensor logistic regression model (BT-LR).

\begin{algorithm}[h!]
\caption{MCMC steps for BT-SVM}
\label{Alg-SVM}
\begin{algorithmic}[1]
\State Update $\rho_i$ from inverse Gaussian: $\rho_i^{-1} \sim IN(\mu_i, \lambda_i)$,  where $\mu_i = |1-y_i(<X_i,B>+z'\gamma)|^{-1}$ and $\lambda_i= 1/\sigma^2$

\State Update hyperparameters $[\alpha, \Phi, \tau \mid \boldsymbol{B}, \boldsymbol{W}]$ compositionally as $[\alpha \mid \boldsymbol{B}, \boldsymbol{W}][\Phi, \tau \mid \alpha, \boldsymbol{B}, \boldsymbol{W}]$, as described in \cite{guhaniyogi2017bayesian}.

\State Sample $\{\beta_j^{(r)}, \omega_{jr}, \lambda_{jr}\}$ using a back-fitting procedure to produce a sequence of draws from the margin-level conditional distributions across components.

\begin{itemize}
    \item[(a)] Draw $\left[w_{j r}, \lambda_{j r} \mid \boldsymbol{\beta}_j^{(r)}, \phi_r, \tau\right]=\left[w_{j r} \mid \lambda_{j r}, \boldsymbol{\beta}_j^{(r)}, \phi_r, \tau\right]\left[\lambda_{j r} \mid \boldsymbol{\beta}_j^{(r)}, \phi_r, \tau\right]$.

    \begin{enumerate}
        \item Draw $\lambda_{j r} \sim \mathrm{Ga}\left(a_\lambda+p_j, b_\lambda+\left\|\boldsymbol{\beta}_j^{(r)}\right\|_1 / \sqrt{\phi_r \tau}\right)$;
        \item Draw $w_{j r, k} \sim \operatorname{giG}\left(\frac{1}{2}, \lambda_{j r}^2, \beta_{j, k}^{2(r)} /\left(\phi_r \tau\right)\right)$ independently for $1 \leq k \leq p_j$
    \end{enumerate}

    \item[(b)] Draw $\beta_j^{(r)}$ from multivariate normal distribution: $\beta_j^{(r)} \sim N(\mu_{jr}, \Sigma_{jr})$, where 
$\mu_{jr} = \frac{\Sigma_{jr}H_j^{(r)}\tilde{\mathbf{y}}}{\sigma^2}$ and $\Sigma_{jr}={(\frac{H_j^{(r)T}H_j^{(r)}}{\sigma^2} + \frac{W_{jr}^{-1}}{\phi_r\tau})}^{-1}$, with
    $$
    h_{i, j, k}^{(r)}=\sum_{d_1=1, \ldots, d_D=1}^{p_1, \ldots, p_D} I\left(d_j=k\right) x_{d_1, \ldots, d_D}\left(\prod_{l \neq j} \beta_{l, i_l}^{(r)}\right),
    $$
    $$
    \boldsymbol{H}_{i, j}^{(r)}=\left(h_{i, j, 1}^{(r)}/ \sqrt{\rho_i}, \ldots, h_{i, j, p_j}^{(r)}/\sqrt{\rho_i}\right)^{\prime},
    $$
    $$
    \tilde{y}_i = \frac{y_i}{\sqrt{\rho_i}}(\rho_i+1-y_i(z_i'\gamma+\sum_{l\neq r} <X_i, B_l>))
    $$
 \end{itemize}

\State Update $\gamma$ from a conjugate normal conditional distribution $\gamma \sim N(\mu_{\gamma}, \Sigma_{\gamma})$
where $\mu_{\gamma} = \Sigma_{\gamma}Z^T (\tilde{y}y/\rho)$ and $\Sigma_{\gamma} = (G^T G/\sigma^2 +\Sigma_{0\gamma})^{-1}$, with
$\boldsymbol{G}_{i, p_z} = Z_{i,p_z}/\sqrt{\rho_i}$ and $\tilde{y_i}=\rho_i+1-y_i<X_i,B>$.
\end{algorithmic}
\end{algorithm}

\begin{algorithm}
\caption{MCMC steps for BT-LR}
\label{Alg-LR}
\begin{algorithmic}[1]

\State Update $\omega_i$ from Polya-gamma distribution: $\omega_i  \sim \mathrm{PG}(1, <X_i, \boldsymbol{B}>+ z_i^T \gamma)$

\State Update hyperparameters $[\alpha, \Phi, \tau \mid \boldsymbol{B}, \boldsymbol{W}]$ compositionally as $[\alpha \mid \boldsymbol{B}, \boldsymbol{W}][\Phi, \tau \mid \alpha, \boldsymbol{B}, \boldsymbol{W}]$, as described in \cite{guhaniyogi2017bayesian}.

\State Sample $\{\beta_j^{(r)}, \omega_{jr}, \lambda_{jr}\}$ using a back-fitting procedure to produce a sequence of draws from the margin-level conditional distributions across components.

\begin{itemize}
    \item[(a)] Draw $\left[w_{j r}, \lambda_{j r} \mid \boldsymbol{\beta}_j^{(r)}, \phi_r, \tau\right]=\left[w_{j r} \mid \lambda_{j r}, \boldsymbol{\beta}_j^{(r)}, \phi_r, \tau\right]\left[\lambda_{j r} \mid \boldsymbol{\beta}_j^{(r)}, \phi_r, \tau\right]$.

    \begin{enumerate}
        \item Draw $\lambda_{j r} \sim \mathrm{Ga}\left(a_\lambda+p_j, b_\lambda+\left\|\boldsymbol{\beta}_j^{(r)}\right\|_1 / \sqrt{\phi_r \tau}\right)$;
        \item Draw $w_{j r, k} \sim \operatorname{giG}\left(\frac{1}{2}, \lambda_{j r}^2, \beta_{j, k}^{2(r)} /\left(\phi_r \tau\right)\right)$ independently for $1 \leq k \leq p_j$
    \end{enumerate}

    \item[(b)] Draw $\beta_j^{(r)}$ from multivariate normal distribution:  $\beta_j^{(r)} \sim N(\mu_{jr}, \Sigma_{jr})$, with
    $\mu_{jr} =\Sigma_{jr}(\Omega (H_j^{(r)})^T \Tilde{y})$ and 
    $\Sigma_{jr} ={((H_j^{(r)})^T \Omega H_j^{(r)}+ W_{jr}^{-1}/(\phi_r\tau))}^{-1},$
where $\Tilde{y} = \kappa/ \omega$, $\kappa=\left(y_1-N_1 / 2, \ldots, y_n-N_n / 2\right)$, $N_1=\ldots=N_n=1$, and $\Omega$ a diagonal matrix with diagonal elements $\omega_i$'s.
\end{itemize}

\State Update $\gamma$ from a conjugate normal conditional distribution $\gamma \sim N(\mu_{\gamma}, \Sigma_{\gamma})$, where $\mu_{\gamma} = \Sigma_{\gamma}Z^T (\tilde{y}\omega)$ and $\Sigma_{\gamma} = (G^T G +\Sigma_{0\gamma})^{-1}$, with
${G}_{i, p_z} = Z_{i,p_z}*\sqrt{\omega_i}$ and $\tilde{y_i}=\kappa_i / \omega_i - <X_i,B>$.
\end{algorithmic}
\end{algorithm}

\section{Simulation Study}\label{Simulation}

\subsection{Data Generation}\label{sec:datagen} 

We illustrate the performance of our methods using several simulation settings and perform comparisons with competitive approaches, based on various types of generated data sets involving several types of functional signals and with data generated from SVM and logistic loss functions. We considered four different types of signals for the tensor coefficient $\boldsymbol{B}$ to generate the binary outcome, as defined below.

{\noindent \underline{Scenario 1:}} In this setting, the tensor $\boldsymbol{B}$ is constructed from rank-\textit{R} PARAFAC decomposition with rank $R_0 = 3$ and dimension $p = c(48, 48)$. Each beta margin $\beta_j^{(r)}$ is generated from the independent binomial distribution $Binomial(2, 0.2)$. After the construction of tensor, we set the maximal value of the tensor $\boldsymbol{B}$ cells to be 1. 
    
{\noindent \underline{Scenario 2:}} The tensor image is simulated by a rank-\textit{R} PARAFAC decomposition with rank $R_0 = 3$.
Here, instead of generating the tensor margin from a known distribution, we manually set up each value of $\beta_j^{(r)}$.
  
{\noindent \underline{Scenario 3:}} Instead of generating the 2D tensor images from a PARAFAC decomposition, the tensor coefficient $\boldsymbol{B}$ is set to be 1 for a rectangular area and 0 otherwise. The non-zero elements cover approximately 30 percent of the area.
    
{\noindent \underline{Scenario 4:}}  The tensor coefficient $\boldsymbol{B}$ is set to be 1 for the circular area and 0 otherwise. The non-zero elements cover approximately 10 percent of the area.

The top panel in Figure \ref{Simulation-figure-proposed} shows the true 2D tensor images, for the different scenarios. For each setting, we generated the tensor covariates $\boldsymbol{X}$ from standard normal distribution $N(0,1)$. For simplicity, we did not include non-tensor covariates in our simulation settings, i.e. we assumed the true $\gamma = (0,\ldots,0)'$. Finally for each scenario, the binary outcome $Y$ was generated according to both the SVM and logistic loss function as follows.  Denoting the linear prediction as  $\psi = <\boldsymbol{X}_i,\boldsymbol{B}>$, the binary outcome is generated as $Y_i = 1$ if $\psi>0$ and $Y_i = -1$ otherwise, under the SVM loss, and from a Bernoulli distribution with probability $p = 1/(1+exp(-\psi)$ under a logistic loss. Therefore a total of 8 scenarios are considered in our simulation set-up, and for each of these scenarios 10 replicates were generated.

\subsection{Parameter Settings}
We choose suitable values of hyperparameters in the prior distributions that yield good overall performance. For example, we set the parameters of the hyperprior on the global scale $\tau$ to $a_\tau = 1$ and $b_\tau = \alpha R^{(1/D)}$, where $R$ is the rank in the assumed PARAFAC decomposition, and set $\alpha_1 = \ldots = \alpha_R = 1/R$. For the common rate parameter $\lambda_{jr}$, we set $a_{\lambda} = 3$ and $b_\lambda=\sqrt[2 D]{a_\lambda}$. Note that under the SVM loss, the scaling parameter $\sigma^2$ is a fixed parameter that can be manually adjusted for maximal model performance. Several values of the tuning parameter $\sigma^2$ from 0.1 through 10 are tested and we choose $\sigma^2 = 6$. In order to decide the rank of the fitted model, we fit the proposed model using ranks 2-5 and choose the rank that minimizes the Deviance Information Criterion (DIC) scores. DIC measures the goodness-of-fit of a set of Bayesian hierarchical models adjusting for model complexity in a manner that penalizes more complex models.

\subsection{Performance Evaluation} 
We report estimation accuracy in terms of relative error(RE), Root Mean Squared Error (RMSE) and correlation coefficient for point estimation of cell-level tensor coefficients. We also illustrate the classification accuracy by calculating misclassification error and the F1 score. Feature selection performances are evaluated by Sensitivity, Specificity, F1 score, and Matthews correlation coefficient (MCC). These metrics are defined as follows. Let $\theta_j,~j = 1,\ldots, J$, be the vectorized tensor coefficients, with $J=\prod_{k=1}^D p_k$ the total number of cells in the tensor coefficient $\boldsymbol{B}$. Further, define the following terms related to classification performance under the SVM classifier: (a) TP is the true positive, i.e. the number of predictions where the classifier correctly predicts the positive class as positive; (b) FP is the false positive, i.e. the number of predictions where the classifier incorrectly predicts the negative class as positive; (c) TN is the true negative, i.e. the number of predictions where the classifier correctly predicts a negative class as negative; and (d) FN is a false negative, i.e. the number of predictions where the classifier incorrectly classifies a positive class as negative. For logistic classification, the above definitions also hold, but after replacing the negative class with zero class. The above definitions of TP/FP/TN/FN can also be adopted for feature selection performance such that the positive class corresponds to non-zero coefficients, and the negative/zero class refers to absent or zero coefficients.

\vskip 12pt

\noindent \underline{Metrics for evaluating coefficient estimation performance:} These metrics include: 
(i) Relative error of $\theta$, denoted as $RE(\theta) = 
     \frac{\sum_{i=1}^{J}|\hat{\theta_i} - \theta_i|}{\sum_{i=1}^{J}|\theta_i|}$, that measures the scaled absolute deviations between the true and estimated parameters,  with $\hat{\theta_i}$ and $\theta_i$ the estimated and true coefficients, respectively. We note that the values of relative error can be larger than 1, since the discrepancy between estimated and true values of tensor cells can be greater than zero when a large portion of the true values is exactly 0. 
     (ii) Root-mean-square error of $\theta$, denoted as $RMSE(\theta) = 
    \sqrt{\frac{\sum_{i=1}^{J} (\hat{\theta_i} - \theta_i)^2}{J}}$, that provides another measure of estimation accuracy; 
    (iii) correlation coefficient between the true and estimated coefficients.

  \vskip 12pt  
\noindent \underline{Metrics for evaluating classification performance:} These metrics include:   
(i) Mis-classification rate, defined as  $\frac{(FP+FN)}{TP+TN+FP+FN}$; and 
(ii) F1-score, defined as the harmonic mean between precision (i.e. TP/(TP+FP)) and recall or sensitivity (TP/(TP+FN)). The expression of F1-score is given as  $\frac{TP}{TP + (FP+FN)/2}$.

 \vskip 12pt    
{\noindent \underline{Metrics for evaluating feature selection performance:} These metrics include: (i) Sensitivity = $\frac{TP}{TP+FN}$; (ii) Specificity = $\frac{TN}{TN+FP}$; and (iii) MCC = $\frac{TP\times TN-FP \times FN}{\sqrt{(TP+FP)(TP+FN)(TN+FP)(TN+FN)}}$.

\vskip 12pt

Results reported below were obtained by randomly splitting the data into training and test sets in the ratio 70:30. The metrics for point estimation and feature selection performances were calculated using the training set data, while the metrics for classification performance were calculated based on the test set. We report below averaged values of the selected metrics across 10 replicates. Two state-of-the-art classification methods are used as competing methods. The first is a penalized logistic regression model with lasso penalty, which is available in the R package \texttt{glmnet} \citep{friedman2010regularization}. The second competing method is the L1-norm SVM model from the R package \texttt{penalizedSVM} \citep{becker2009penalizedsvm,Bradley1998FeatureSV}. Both methods use a vectorization approach, where the tensor covariates are vectorized into scalar variables and then fit into the statistical models. Therefore they do not respect the spatial information in the image. Additionally, we use a grid search algorithm and cross-validation to select the best tuning parameters prior to model fitting.

\subsection{Results}

We ran MCMC chains for 3,000 iterations, with 1,000 burn-in iterations. The computation time varies depending on the rank and is expected to increase with higher ranks. It took around 19 minutes to run a single MCMC chain with rank 2, and around 36 minutes with rank 4. A Geweke diagnostic \citep{geweke1991evaluating} was applied to examine for signs of non-convergence parameters. We obtain the z-score from Geweke for each element of the coefficient matrix $\boldsymbol{B}$. For the proposed Bayesian tensor SVM model (BT-SVM), we observe the z-scores lie in the range $(-1.96, 1.96)$ for 91 percent of the coefficient matrix elements. For the proposed Bayesian tensor logistic regression model (BT-LR), the z-scores lie in the range $(-1.96, 1.96)$ for 79 percent, indicating that most chains have reached ergodicity.

\begin{table*}[hbt!]
\caption{Point Estimation and Out-of-sample Classification results for the four 2D tensor images portrayed in Figure \ref{Simulation-figure-proposed} top panel; Y generated from SVM loss}\label{sim-estimation-svm}
\begin{tabular*}{\textwidth}{@{\extracolsep\fill}llcccccc}
\toprule%
Scenarios & Methods & RE & RMSE & Corr.Coef. & Mis. Class. & F1-score \\

\midrule
Scenario 1  & LR w/ lasso & 1.000 & 0.558 & 0.071 & 0.54 & 0.147\\
& L1norm-SVM &1.110   &0.687  &0.027  &0.513  &0.522    \\
&{BT-SVM} & 0.986  & 0.489 & 0.442 & 0.260  & 0.764 \\
& {BT-LR} & 1.055 & 0.504 & 0.383 & 0.342 & 0.689  \\
\midrule
Scenario 2 &LR w/ Lasso &1.000 & 0.382 & 0.055 & 0.46 & 0 \\
&L1norm-SVM & 1.226 & 0.533 &0.092  & 0.48 & 0.502  \\
&BT-SVM & 0.853  & 0.246 & 0.874 & 0.149  & 0.837 \\
&BT-LR & 1.006 &0.277 & 0.794 & 0.204 & 0.778\\
\midrule
Scenario 3 &LR w/ Lasso & 1.001 &0.541 &0.074 & 0.507&0.191 \\
&L1norm-SVM & 1.032 & 0.537 & 0.109 & 0.533  & 0.512  \\
&BT-SVM & 0.836  & 0.412 & 0.810 & 0.197  & 0.823 \\
&BT-LR & 0.909 & 0.439 & 0.733 & 0.238 & 0.781 \\
\midrule
Scenario 4&LR w/ Lasso &1.005 & 0.330 & 0.122 & 0.533 & 0.2\\
&L1norm-SVM & 1.335 & 0.535 & 0.053 & 0.44  & 0.565 \\
&BT-SVM & 1.013  & 0.225 & 0.773 & 0.200  & 0.819\\
&BT-LR & 1.447  & 0.262 & 0.614 & 0.272 & 0.752\\
\botrule
\end{tabular*}
\end{table*}

\begin{table*}[hbt!]
\caption{Feature selection results for the four 2D tensor images portrayed in Figure \ref{Simulation-figure-proposed} top panel; Y generated from SVM loss}\label{sim-feature-selection-svm}
\begin{tabular*}{\textwidth}{@{\extracolsep\fill}llcccccc}
\toprule%
Scenarios & Methods & Sens. & Spec. & F1-score & MCC  \\

\midrule
Scenario 1  & LR w/ lasso & 0.031 & 0.984 & 0.057 & 0.046 \\
& L1norm-SVM & 0.058  &0.950  &0.1  & 0.017 \\
&{BT-SVM} & 0.070  & 0.999 & 0.130 & 0.213 \\
& {BT-LR} &  0.214  & 0.938 & 0.313 & 0.225 \\
\midrule
Scenario 2 &LR w/ Lasso & 0.003 & 1 &0.007  &0.055  \\
&L1norm-SVM & 0.021 & 0.985 &0.037  & 0.016   \\
&BT-SVM & 0.628  & 1 & 0.771 & 0.772  \\
&BT-LR & 0.821 & 0.978 & 0.835 & 0.814\\
\midrule
Scenario 3 &LR w/ Lasso & 0.025 & 0.981 & 0.047 & 0.019  \\
&L1norm-SVM & 0.019 & 0.988 & 0.037 & 0.027    \\
&BT-SVM & 0.192 & 1 & 0.32 &  0.377  \\
&BT-LR & 0.517 & 0.987 & 0.668 & 0.626  \\
\midrule
Scenario 4&LR w/ Lasso & 0.094 & 0.974 & 0.145 & 0.122  \\
&L1norm-SVM & 0.020 & 0.984 & 0.034 & 0.009   \\
&BT-SVM & 0.368 & 0.998 & 0.530 & 0.573  \\
&BT-LR & 0.519 & 0.934 & 0.552 & 0.522 \\
\botrule
\end{tabular*}
\end{table*}

\begin{table*}[hbt!]
\caption{Point Estimation and Out-of-sample Classification results for the four 2D tensor images portrayed in Figure \ref{Simulation-figure-proposed} top panel; Y generated from logistic regression loss}\label{sim-estimation-logistic}
\begin{tabular*}{\textwidth}{@{\extracolsep\fill}llcccccc}
\toprule%
Scenarios & Methods & RE & RMSE & Corr.Coef. & Mis. Class. & F1-score \\

\midrule
Scenario 1  & LR w/ lasso & 1 & 0.557 & 0.076 & 0.56 & 0.125\\
& L1norm-SVM &1.017 &0.562  &0.054  &0.48  &0.55    \\
&{BT-SVM} & 0.991  & 0.493 & 0.495 & 0.274  & 0.745 \\
& {BT-LR} & 1.010 &0.490  & 0.445 & 0.247 & 0.767  \\
\midrule
Scenario 2 &LR w/ Lasso &1.000    &0.382  &0.085  &0.46 & 0 \\
&L1norm-SVM & 1.034 & 0.383 & 0.053 & 0.413 & 0.537 \\
&BT-SVM & 0.875  & 0.245 &0.859 & 0.179  & 0.804 \\
&BT-LR &1.196  &0.260  & 0.732 & 0.252 & 0.734\\
\midrule
Scenario 3 &LR w/ Lasso &  1.000  & 0.542 & 0.071 & 0.506 & 0.380 \\
&L1norm-SVM & 1.068 & 0.587 & 0.041 & 0.433  &  0.586 \\
&BT-SVM & 0.836 & 0.412 & 0.804 & 0.178  & 0.842 \\
&BT-LR & 0.964 & 0.421 & 0.618 & 0.253 & 0.770 \\
\midrule
Scenario 4&LR w/ Lasso &  1.002  & 0.331 & 0.131 & 0.467 & 0.557 \\
&L1norm-SVM & 1.361 & 0.542 & 0.058 & 0.467 & 0.557 \\
&BT-SVM & 1.017 & 0.226 & 0.766 & 0.203 & 0.813\\
&BT-LR & 1.202 & 0.255 & 0.658 & 0.227 & 0.788\\
\botrule
\end{tabular*}
\end{table*}

\begin{table*}[hbt!]
\caption{Feature selection results for the four 2D tensor images portrayed in Figure \ref{Simulation-figure-proposed} top panel; Y generated from logistic regression loss}\label{sim-feature-selection-logistic}
\begin{tabular*}{\textwidth}{@{\extracolsep\fill}llcccccc}
\toprule%
Scenarios & Methods & Sens. & Spec. & F1-score & MCC  \\

\midrule
Scenario 1  & LR w/ lasso & 0.036 & 0.979 & 0.067 & 0.045 \\
& L1norm-SVM & 0.023  &0.982  &0.044  & 0.020 \\
&{BT-SVM} &0.007 &0.999 &0.014 & 0.061 \\
& {BT-LR} & 0.343 & 0.913 & 0.446 & 0.318 \\
\midrule
Scenario 2 &LR w/ Lasso & 0.017 & 0.999 & 0.034 & 0.098 \\
&L1norm-SVM &0.042 &0.978 & 0.070 & 0.042\\
&BT-SVM &0.590 &0.999 &0.736 &0.742  \\
&BT-LR & 0.801 & 0.948 & 0.764 & 0.736 \\
\midrule
Scenario 3 &LR w/ Lasso &0.024 & 0.984 & 0.044 & 0.026 \\
&L1norm-SVM & 0.010&0.981 &0.020 & -0.031 \\
&BT-SVM &0.180 & 1 &0.300 &0.361 \\
&BT-LR & 0.475 & 0.926 & 0.573 & 0.483  \\
\midrule
Scenario 4&LR w/ Lasso & 0.086 & 0.972 & 0.133 & 0.102\\
&L1norm-SVM & 0.040 & 0.983 & 0.067 & 0.051 \\
&BT-SVM & 0.409& 0.999& 0.577& 0.612\\
&BT-LR & 0.609 & 0.965 & 0.645 & 0.610\\
\botrule
\end{tabular*}
\end{table*}

We report results for estimation and classification accuracy, and feature selection in Tables \ref{sim-estimation-svm}, \ref{sim-feature-selection-svm}, \ref{sim-estimation-logistic}, \ref{sim-feature-selection-logistic}. Specifically, Tables \ref{sim-estimation-svm} and \ref{sim-feature-selection-svm} reflect results across all 4 scenarios when the binary outcome Y is generated from a SVM loss, while Tables \ref{sim-estimation-logistic} and \ref{sim-feature-selection-logistic} reflect the logistic type of loss. These results demonstrate that both the proposed methods (BT-SVM and BT-LR) consistently outperform competing penalized methods in terms of coefficient estimation, feature selection, and classification performance across all scenarios. When the binary outcome data is generated from SVM loss, the BT-SVM approach has superior coefficient estimation (as evident from lower RE/RMSE and higher correlation coefficient in Table \ref{sim-estimation-svm}) and improved classification accuracy (as evident from lower misclassification rate and higher F1-score in Table \ref{sim-estimation-svm}). Even when the data is generated from a logistic loss, the same trends hold, with the exception of Scenario 1 where the proposed BT-LR approach reports improved classification accuracy and comparable coefficient estimation, as evident from the results in Table \ref{sim-estimation-logistic}. 

Moreover, the proposed model with SVM loss (BT-SVM) generally performs worse than the corresponding model with logistic loss (BT-LR) in terms of feature selection, as evident from the results in Tables \ref{sim-feature-selection-svm} and \ref{sim-feature-selection-logistic}. In particular, the BT-LR approach almost always has considerably higher sensitivity compared to the BT-SVM model while having comparable or slightly lower specificity, even when the outcome data is generated under the SVM loss. This results in the BT-LR approach consistently having higher F1-score and higher or comparable MCC values for feature selection, when compared with the BT-SVM approach, even when the outcome data is generated under the SVM loss. 

\begin{figure}[hbt]%
\centering
\includegraphics[width=1\textwidth]{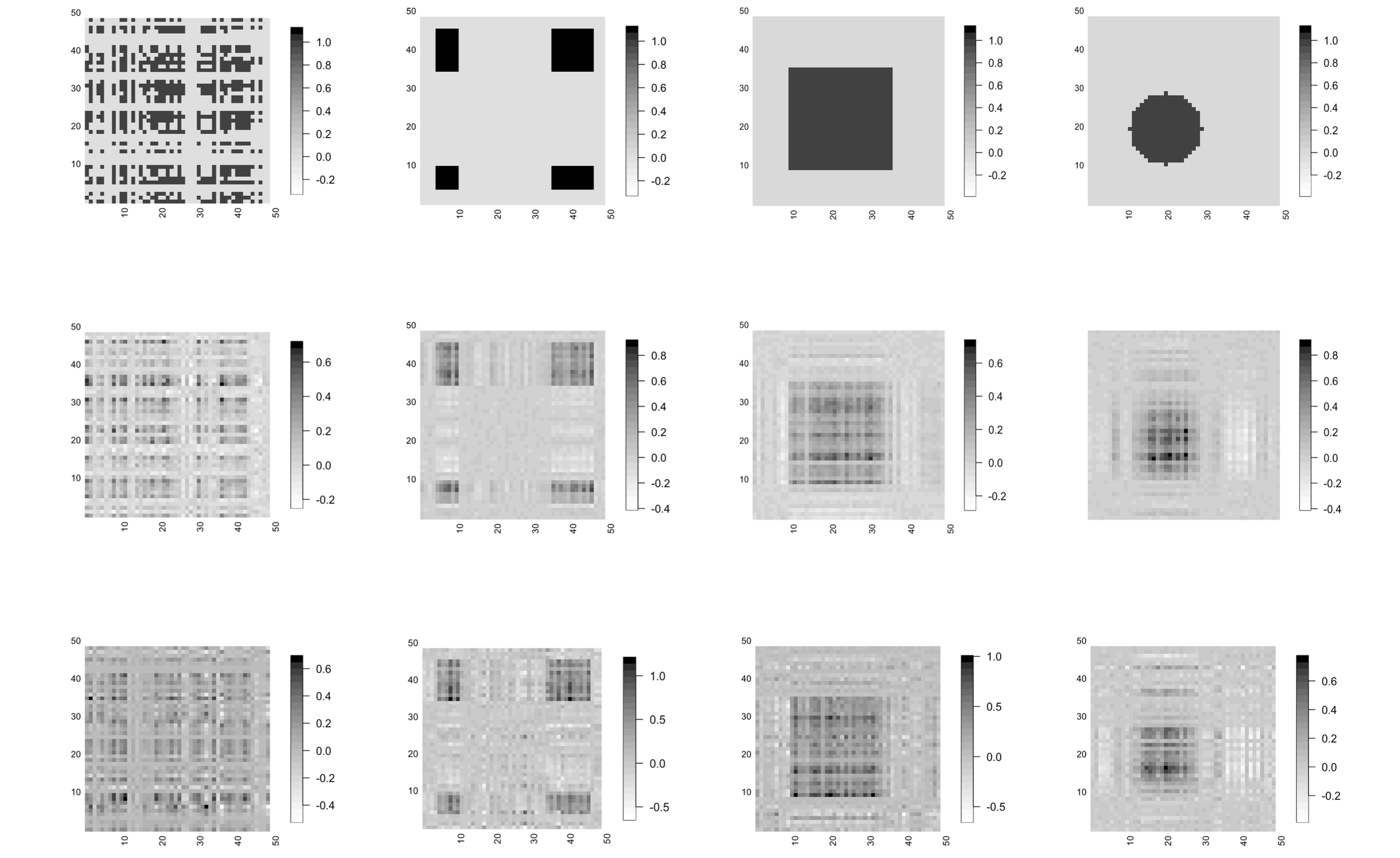}
\caption{Row 1 from left: Simulated data with 48$\times$48 2D tensor images from Scenario 1, Scenario 2, Scenario 3, and Scenario 4. Row 2: Recovered images for the 48$\times$48 2D tensor images using BT-SVM corresponding to 4 scenarios in row 1. Row 3: Recovered images for the 48$\times$48 2D tensor images using BT-LR corresponding to 4 scenarios in row 1.}
        \label{Simulation-figure-proposed}
\end{figure}

Combining the above discussions, the BT-SVM model appears to generally show improved coefficient estimation and classification performance over its counterpart with logistic loss, regardless of which type of loss function was used to generate the outcome data. On the contrary, the BT-LR model appears to have improved sensitivity and comparable specificity compared to its counterpart with SVM loss that translates to improved feature selection, regardless of which loss function is used to generate the outcome data. Figure \ref{Simulation-figure-proposed} presents the coefficient matrix cell estimation using the proposed methods BT-SVM and BT-LR. From this Figure, it is evident that the proposed method is able to broadly recover the shapes of the 2D tensor $\boldsymbol{B}$ regardless of whether the underlying signal is generated using the PARAFAC decomposition or not.

\begin{figure}[hbt]%
\centering
\includegraphics[width=1\textwidth]{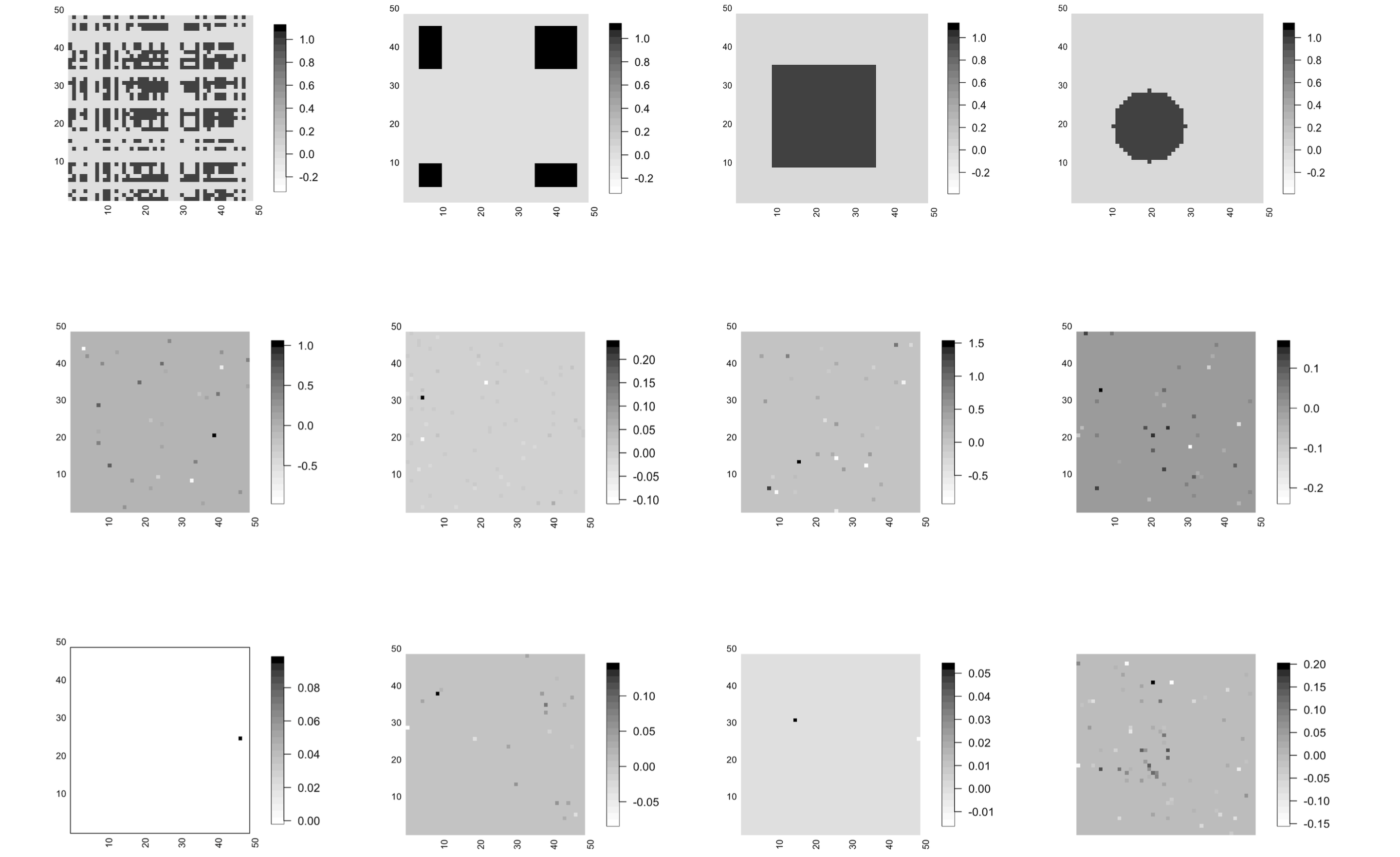}
\caption{Row 1 from left: Simulated data with 48$\times$48 2D tensor images from Scenario 1, Scenario 2, Scenario 3, and Scenario 4. Row 2: Recovered images for the 48$\times$48 2D tensor images using competing method L1norm-SVM for Scenario 1, Scenario 2, Scenario 3, and Scenario 4. Row 3: Recovered images for the 48$\times$48 2D tensor images using logistic regression with LASSO penalty.}
        \label{Simulation-figure-competing}
\end{figure}

In contrast, the penalized competing methods (logistic regression with LASSO penalty and L1norm-SVM) illustrate poor estimation, feature selection, and out-of-sample classification performance. This is evident from the results in Tables \ref{sim-feature-selection-svm}, \ref{sim-feature-selection-logistic}, where the competing approaches report notably low sensitivity, almost approaching zero. This low sensitivity reflects their inability to detect the true signals, resulting in poor coefficient estimation performance as evident from Tables \ref{sim-estimation-svm}, \ref{sim-estimation-logistic}, where the correlation coefficients between the true and estimated coefficients are often close to zero. For a more detailed perspective, Figure \ref{Simulation-figure-competing} visualizes the estimated coefficients under two competing methods. The Figure reveals spatially disparate non-zero coefficients, indicating that the absence of spatial smoothing hampers the accurate estimation of true signals by these competing methods. Ultimately, the substandard feature selection performance contributes to considerably inferior classification results, as demonstrated in Tables  \ref{sim-estimation-svm},  \ref{sim-estimation-logistic}.

\subsection{Sensitivity Analysis}

\cite{guhaniyogi2017bayesian} set up a series of default values for the prior hyper-parameters for the Bayesian tensor regression model. Specifically, hyperparameter of the Dirichlet component $\left(\alpha_1, \ldots, \alpha_R \right) =\alpha =  (1/R, ..., 1/R)$, $a_\lambda = 3$,  $b_\lambda=\sqrt[2 D]{a_\lambda}$, $a_\tau = \sum_{i=1}^R \alpha_i$, and $b_\tau = \alpha R^{(1/D)} $ are set as defaulted values to control the cell-level variance on tensor $\boldsymbol{B}$. We conducted a prior hyperparameter sensitivity analysis by tuning each hyperparameter while fixing other hyperparameters. Table \ref{hyperparameter-sensitivity} displays the cell-level RMSE values of the tensor coefficient matrix $\boldsymbol{B}$ of dimension $D=2$ and parafac rank-\textit{R}, where for BT-SVM model,  $\boldsymbol{B}$ is generated from Scenario 4; and for BT-LR model, $\boldsymbol{B}$ is generated from Scenario 3. The overall results reveal no strong sensitivity to the hyper-parameter choices under the selected ranges that point to a robust performance.

\begin{table}[hbt!]
\caption{Prior hyperparameter sensitivity analysis}\label{hyperparameter-sensitivity}
\begin{tabular}{@{}lllllll@{}}
\toprule
 & Hyperparameters&RMSE&Hyperparameters&RMSE&Hyperparameters&RMSE \\
\midrule
 \multirow{2}{*}{BT-SVM} & $\alpha$ = 1/9  & 0.224 & $a_{\lambda}$ = 3 & 0.225 &$a_{\tau}$ = 1/3 & 0.225 \\
 &$\alpha$= 1/6 &0.228 &$a_{\lambda}$ = 5 & 0.235 & $a_{\tau}$ = 1/2 & 0.225 \\
 & $\alpha$ = 1/3& 0.226 &$a_{\lambda}$ = 7 & 0.227 &$a_{\tau}$ = 1 & 0.226  \\
 & $\alpha =  3^{(-0.1)}$ & 0.227 &$a_{\lambda}$ = 10 &0.230&$a_{\tau}$ = 2 &0.231 \\
\midrule
 \multirow{2}{*}{BT-LR} & $\alpha$ = 1/9  & 0.418 & $a_{\lambda}$ = 3 & 0.413 &$a_{\tau}$ = 1/3 & 0.428 \\
 &$\alpha$= 1/6 &0.422 &$a_{\lambda}$ = 5 & 0.421 & $a_{\tau}$ = 1/2 & 0.416 \\
 & $\alpha$ = 1/3& 0.419 &$a_{\lambda}$ = 7 & 0.413 &$a_{\tau}$ = 1 & 0.416  \\
 & $\alpha =  3^{(-0.1)}$ & 0.421 &$a_{\lambda}$ = 10 &0.415&$a_{\tau}$ = 2 &0.420 \\
\botrule
\end{tabular}
\end{table}

\section{ADNI Data Analysis}\label{ADNI-section}

\subsection{Data Source and Pre-processing}
This study utilized the data obtained from the Alzheimer’s Disease Neuroimaging Initiative (ADNI), which is a longitudinal multicenter study launched in 2004 for the early detection and tracking of Alzheimer’s disease (AD). ADNI researchers collect multiple data types such as clinical, behavioral and genetic data, along with neuroimaging measurements such as magnetic resonance imaging (MRI), positron emission tomography (PET), and biospecimens. We use data from the ADNI 1 study collected at baseline, consisting of T1-weighted MRI scans, cognitive measurements in terms of Mini Mental State examination (MMSE), and basic demographic data (age, gender, year of education, APOE status) of 818 subjects. A more detailed description of the ADNI data is provided in Table \ref{ADNI-demographic}.

\begin{table}[hbt!]
\caption{Summary of demographic variables and cognitive measurements under study}\label{ADNI-demographic}
\begin{tabular}{@{}ccc@{}}
\toprule%
& & \textbf{Overall} (N = 818) \\
\midrule
\multirow{2}{*}{\textbf{Age}} & Mean (SD) &  75.17 (6.83)\\
& Median[Min, Max] & 75.45[54.40, 90.90] \\
\midrule
\multirow{2}{*}{\textbf{Years of Education}} & Mean (SD) & 15.53 (3.05) \\
& Median [Min, Max] & 16 [4, 20]\\
\midrule
\multirow{2}{*}{\textbf{Gender}} & Female & 342 (41.8\%) \\
& Male & 476 (58.2\%)\\
\midrule
\multirow{3}{*}{\textbf{APOE4}} & 0 & 418 (51.1\%)\\
& 1 & 312 (31.2\%)\\
& 2 & 88 (10.7\%)\\
\midrule
\multirow{2}{*}{\textbf{MMSE score}} & Mean(SD) & 26.74 (2.67)  \\
& Median[Min, Max] & 27 [18, 30]\\
\midrule
\multirow{2}{*}{\textbf{log(ICV score)}} & Mean (SD) & 14.25 (0.11) \\
& Median [Min, Max] & 24.25 [13.92, 14.56]\\
\botrule
\end{tabular}
\end{table}

The T1-weighted MRI images underwent processing through the Advance Normalization Tools (ANTs) registration pipeline \citep{tustison2014large}, where all images were registered to a template image to ensure consistent normalization of brain locations across participants. The population-based template was constructed based on data from 52 normal control participants in ADNI 1, originally from the ANTs group \citep{tustison2019longitudinal}. Notably, the ANTs pipeline includes the N4 bias correction step, addressing intensity discordance to inherently standardize intensity across samples \citep{tustison2010n4itk}. It also employs a symmetric diffeomorphic image registration algorithm for spatial normalization, aligning each T1 image with a brain image template to facilitate spatial comparability \citep{avants2008symmetric}. Subsequently, the pipeline utilized the processed brain images, estimated brain masks, and template tissue labels for 6-tissue Atropos segmentation, generating tissue masks for cerebrospinal fluid (CSF), gray matter (GM), white matter (WM), deep gray matter (DGM), brain stem, and cerebellum. Finally, cortical thickness measurements were derived using the DiReCT algorithm. The 3-D cortical thickness image was further downsampled to a dimension of $48 \times 48 \times 48$  and divided into 48 2-D axial slices of dimension $48 \times 48$, and a subset of these 2D slices were used for our analysis. The downsampling step reduces the dimension of the image, and also somewhat alleviates the sparsity of the cortical thickness maps by consolidating adjacent voxels and presenting the average cortical thickness. A reduction of sparsity proves beneficial to fitting our Bayesian tensor model, and the same should hold of other commonly used statistical models.

\subsection{Analysis Outline}

We apply the proposed approaches to perform various classification tasks using data from the Alzheimer’s Disease Neuroimaging Initiative (ADNI) study. Demographic information of age, gender, years of education, APOE4 allele (0,1,2) and intracranial volume (ICV) are incorporated as scalar covariates, while 2D cortical thickness slices (derived from the T1 weighted MRI scans) are used as tensor covariates. Since the different 2D brain slices are expected to contain varying amount of the brain cortical regions, it is important to choose these slices carefully. In particular, we would like to choose 2D slices such that they contain at least a certain portion of the brain cortex, in order to contain enough information to perform classification. Therefore, we present the analysis results from 7 different axial slices (we will denote them slices 19 to 25), each of which has cortical brain regions that cover at least 65\% of the slice. We evaluate the classification performance of the proposed approaches and compare with the penalized logistic regression with lasso and L1-norm SVM as described in Section \ref{Simulation}.

In particular, we assess the ability of the proposed classifiers to differentiate between different types of disease phenotypes using the 2D imaging slices along with demographic information. In particular, we perform the following types of classification tasks corresponding to disease phenotypes: (i) normal control vs. AD patients; (ii) normal control vs. MCI patients; (iii) MCI vs. AD patients. In addition to the above tasks, we also perform (iv) gender classification (males vs females); and (v) intelligence level classification based on high and low levels of MMSE scores. MMSE is commonly used clinically for checking cognitive impairment with a low value indicative of a cognitive decline. We stratify individuals into high vs low MMSE categories, depending on whether their MMSE scores are above the 70th percentile or below the 30th percentile of the MMSE distribution. The corresponding sample sizes in the high and low intelligence categories were 280 and 249 respectively. For each classification task based on a given 2D slice, the data is randomly split into training and test splits in the ratio 70:30, and 10 such splits are considered. The results are averaged over these 10 replicates and reported in Table \ref{ADNI-results}.

\subsection{Results}

\begin{table}[hbt!]
\caption{Classification on Gender, Disease phenotype, and High/Low MMSE scores with Demographic covariates.}\label{ADNI-results}
\begin{tabular*}{\textwidth}{@{\extracolsep\fill}lcccccccc}
\toprule%
& \multicolumn{2}{@{}c@{}}{BT-SVM} & \multicolumn{2}{@{}c@{}}{BT-LR}& \multicolumn{2}{@{}c@{}}{LR-lasso} & \multicolumn{2}{@{}c@{}}{L1norm-SVM}\\\cmidrule{2-3}\cmidrule{4-5}\cmidrule{6-7}\cmidrule{8-9}%
 & Mis. C.\footnotemark[1] & F1-score & Mis. C. & F1-score  & Mis. C. & F1-score & Mis. C. & F1-score \\
\midrule
 & \multicolumn{8}{@{}c@{}}{ Female vs. Male }  \\
 \midrule
 slice 19& 0.288&0.764&0.317 &0.728 &0.337 &0.750 &0.422 & 0.636  \\
 slice 20&0.280 &0.762  & 0.292 &0.748  &0.361 &0.729 & 0.398 & 0.654\\
 slice 21 &0.296 &0.752 &0.3 &0.756 &0.357 &0.75 & 0.447 & 0.595\\
 slice 22&0.284 &0.779 &0.333 &0.723 &0.361 &0.729 & 0.504 & 0.550\\
 slice 23 &0.256 &0.789 &0.325 &0.733 &0.373 &0.657 &0.390 &0.684 \\
 slice 24 &0.280 &0.771 &0.325 &0.720 &0.398 &0.611 &0.447 &0.618 \\
 slice 25 &0.256 &0.801 &0.345 &0.719 &0.369 &0.674 &0.455 &0.591 \\
\midrule
 & \multicolumn{8}{@{}c@{}}{ NC. vs. AD }  \\
 \midrule
 slice 19&0.327 &0.669 &0.277 &0.715 &0.388 &0.595 & 0.452 & 0.564\\
 slice 20&0.301 &0.712 &0.325 &0.682 &0.341 &0.644 & 0.365 & 0.671\\
 slice 21&0.341  &0.656  &0.309 &0.697 &0.341&0.638 & 0.404 & 0.564\\
 slice 22&0.278  &0.724  &0.325 &0.655 &0.333 &0.655 & 0.444 & 0.582\\
 slice 23&0.277 &0.720 &0.309 &0.698 &0.293 &0.654 &0.436 &0.444 \\
 slice 24&0.293 &0.689 &0.333 & 0.681 &0.325 &0.601 &0.341 &0.626 \\
 slice 25&0.269 &0.746 &0.309 &0.677 &0.301 &0.660 &0.396 & 0.510\\
 \midrule
 & \multicolumn{8}{@{}c@{}}{ NC. vs MCI }  \\
 \midrule
  slice 19 &0.280 &0.8 &0.312 &0.768  &0.344 &0.732 & 0.414 & 0.686  \\
 slice 20 &0.291 &0.791 &0.380 &0.707 &0.302 &0.751 & 0.418 & 0.691  \\
 slice 21 &0.338 &0.761 &0.365 &0.718 &0.333&0.731 & 0.440 & 0.688 \\
 slice 22 &0.317 &0.771 &0.349 &0.736 &0.333 &0.722 & 0.458 & 0.664  \\
 slice 23&0.322 &0.776 &0.349 &0.766 &0.365 &0.696 &0.455 &0.586 \\
 slice 24 &0.328 &0.760 &0.349 &0.750 &0.360 &0.704 &0.433 &0.620 \\
 slice 25 &0.291 &0.792 &0.354 &0.729 &0.333 &0.720 &0.465 &0.582 \\
\midrule
 & \multicolumn{8}{@{}c@{}}{ AD vs. MCI }  \\
\midrule
 slice 19&0.296  &0.805  &0.355 &0.746  &0.334 &0.763 & 0.395 & 0.690\\
 slice 20& 0.282 & 0.816 &0.305 &0.784 &0.325 &0.577 & 0.420 & 0.495 \\
 slice 21& 0.287 & 0.815 &0.338 &0.781 &0.350&0.759 & 0.412 & 0.691  \\
 slice 22& 0.282 & 0.810 &0.322 &0.753 &0.322 &0.778 & 0.356 & 0.770  \\
 slice 23& 0.271& 0.825&0.344 &0.751 &0.384 &0.723 &0.401 &0.702\\
 slice 24& 0.282& 0.812&0.338 &0.758 &0.350 &0.747 &0.446 &0.663\\
 slice 25&0.291 &0.808 &0.389 &0.721 &0.339 &0.765 &0.463 &0.613\\
\midrule
 & \multicolumn{8}{@{}c@{}}{ High vs Low MMSE scores }  \\
 \midrule
 slice 19& 0.283 & 0.723 &0.327 &0.675  &0.333 &0.693 & 0.496 & 0.606  \\
 slice 20& 0.295 &0.715 &0.358 &0.655 &0.333 &0.674 & 0.471 & 0.534  \\
 slice 21& 0.289  &0.722  &0.352 &0.654 &0.345 &0.667 & 0.484 & 0.549  \\
 slice 22&0.295 &0.710 &0.339 &0.686 &0.377 &0.634 & 0.471 & 0.460  \\
 slice 23&0.295 &0.712 &0.345 &0.645 &0.327 &0.653 &0.421 &0.645 \\
 slice 24&0.289 &0.720 &0.333 &0.675 &0.321 &0.622 &0.459 &0.587 \\
 slice 25&0.301 &0.707 &0.327 &0.662 &0.327 &0.653 &0.471 &0.590 \\
\botrule
\end{tabular*}
\footnotetext[1]{Misclassification Rate.}
\end{table}

Table \ref{ADNI-results} reports the misclassification rate and f1-score for slices 19 through 25. The proposed approach under both loss functions almost always results in better classification accuracy consistently across slices and classification tasks, compared to the penalized approaches. While both the two competing approaches have inferior performance compared to their Bayesian counterparts, the LR-Lasso generally performs slightly better than the L1norm-SVM. Among the three disease phenotype classification tasks, the highest accuracy is achieved for AD vs MCI classification under the BT-SVM approach, with f1-scores greater than 0.8 across all 2D slices. The proposed approaches are also able to perform well for the NC vs MCI classification task with F1-scores consistently greater than 0.75, while the classification performance for NC vs AD is slightly less impressive under all methods. We note that slice 23 provides the highest accuracy for both AD vs MCI and NC vs MCI classification tasks. With regards to the classification of non-imaging phenotypes, the gender classification performance is generally improved compared to intelligence classification, which is not surprising given the considerably lower training sample size for classifying the high versus low intelligence cohorts. 

Clearly, there exist fluctuations in classification performances across the different 2-D slices. This is not surprising because cortical thickness contained in each 2-D slice varies depending on the sections of the slice. Further, different slices represent different brain regions that may have differential effects on classifying phenotypic classes. Overall, these results demonstrate that our proposed Bayesian tensor approach offers consistent improvements for classifying Alzheimer's disease and subjects' demographic information, by leveraging the spatial information in the images and incorporating dimension reduction that potentially avoids overfitting. Moreover, the unsatisfactory performance of the competing methods clearly illustrates the perils of ignoring the spatial correlations in high-dimensional images in classification tasks. This is not surprising, given that Lasso-based approaches are known to be affected by multicollinearity. While it is possible to explore alternate approaches such as principal component regression (PCR), such approaches lead to loss of interpretability that is not desirable in imaging studies due to various reasons including the inability to perform feature selection.

\section{Discussion}\label{Discussion}

In this study, we proposed a Bayesian tensor classification approach based on high-dimensional neuroimaging data and scalar predictors, via data augmentation. The proposed approach essentially extends the literature on Bayesian tensor regression to classification problems that has the ability to perform inference and uncertainty quantification. The tensor structure is particularly suitable for neuroimaging data since it respects the spatial information of imaging voxels while keeping the number of parameters to be estimated at a manageable level via a PARAFAC decomposition. With two data augmentation schemes implemented, we demonstrated the superiority of the proposed method via simulation and data application. In particular, comprehensive simulation studies concretely illustrate the advantages under the proposed approach, with each type of data augmentation having distinct advantages in classification, coefficient estimation and/or feature selection. We applied the proposed method to the ADNI dataset to classify disease phenotype, as well as gender and intelligence levels. Both data augmentation schemes showed consistently higher classification accuracy and improved feature selection than other penalized methods that vectorized the image. This showed the benefits of incorporating the spatial information in the image.

A potential limitation of the study is that we use 2D brain slices instead of using the whole 3D brain image. While using the full image may potentially result in improvements in accuracy, it can also cause computational bottlenecks due to a massive number of model parameters. In future work, we intend to propose more scalable versions of the proposed approach that can be used to incorporate high-dimensional 3D images for classification. We also propose to explore various prior choices for the tensor coefficients in order to enable stronger shrinkage and thicker tails, with a view to improving the feature selection performance. Another potential limitation is that we did not consider potential heterogeneity across samples in our modeling scheme. In future work, we plan to account for the heterogeneity across samples via a mixture of tensors approach that is particularly relevant for AD studies. While we plan to investigate the above issues in future research, we believe that the proposed approach in this article provides a valuable Bayesian classification tool based on imaging covariates that fills an important gap in literature.

\backmatter

\section{Declarations}

\bmhead{Information Sharing Statement}

We will make our code available on Github upon acceptance of the paper. The sample data from ADNI can be accessed by accepting the Data Use Agreement and submitting an online application form at \url{https://adni.loni.usc.edu/data-samples/access-data/}.

\bmhead{Author Contributions}
S.K. designed the research and conceived the analysis plan; R.L. performed model implementation and data analyses; S.K. and M.V. provided overall supervision and direction of the work. R.L., S.K. and M.V. wrote the paper.

\bmhead{Competing Interests}
The authors have no relevant financial or non-financial interests to disclose. The authors have no competing interests to declare that are relevant to the content of this article. All authors certify that they have no affiliations with or involvement in any organization or entity with any financial interest or non-financial interest in the subject matter or materials discussed in this manuscript. The authors have no financial or proprietary interests in any material discussed in this article.

\FloatBarrier

\bibliography{bibliography}

\end{document}